\documentclass[showpacs,twocolumn,twoside,pra,superscriptaddress,preprintnumbers,byrevtex]{revtex4-1}
\usepackage{amsmath, amssymb, bm}
\usepackage{graphicx}
\usepackage[font=footnotesize,caption=false]{subfig}
\usepackage{enumerate}
\usepackage{color}
\usepackage[colorlinks=true,citecolor=blue,linkcolor=blue,urlcolor=blue]{hyperref}

\usepackage{bbm}

\begin{document}

\title{General description of quasi-adiabatic dynamical phenomena near exceptional points}

\author{Thomas J. Milburn}
\email{thomas.milburn@ati.ac.at}
\affiliation{Institute of Atomic and Subatomic Physics, Vienna University of Technology (TU Wien), Stadionallee 2, 1020 Vienna, Austria}
\author{J\"org Doppler}
\affiliation{Institute for Theoretical Physics, Vienna University of Technology (TU Wien), 1040 Vienna, Austria}
\author{Catherine A. Holmes}
\affiliation{University of Queensland, School of Mathematics and Physics, QLD 4072, Australia}
\author{Stefano Portolan}
\affiliation{University of Southampton, Department of Physics and Astronomy, Southampton SO17 1BJ, United Kingdom}
\author{Stefan Rotter}
\affiliation{Institute for Theoretical Physics, Vienna University of Technology (TU Wien), 1040 Vienna, Austria}
\author{Peter Rabl}
\affiliation{Institute of Atomic and Subatomic Physics, Vienna University of Technology (TU Wien), Stadionallee 2, 1020 Vienna, Austria}

\begin{abstract}
The appearance of so-called exceptional points in the complex spectra of
non-Hermitian systems is often associated with phenomena that contradict our
physical intuition. One example of particular interest is the state-exchange
process predicted for an adiabatic encircling of an exceptional point. In this
work we analyse this and related processes for the generic system of two
coupled oscillator modes with loss or gain. We identify a characteristic
system evolution consisting of periods of quasi-stationarity interrupted by
abrupt non-adiabatic transitions, and we present a qualitative and
quantitative description of this switching behaviour by connecting the problem
to the phenomenon of stability loss delay. This approach makes accurate
predictions for the breakdown of the adiabatic theorem as well as the
occurrence of chiral behavior observed previously in this context, and
provides a general framework to model and understand quasi-adiabatic dynamical
effects in non-Hermitian systems.
\end{abstract}

\pacs{}

\preprint{Preprint of \today---not for dissemination}

\maketitle


\section{Introduction}
\label{sec:Introduction}

The quantum adiabatic theorem is a seminal result in the history of quantum
mechanics. Paraphrasing Born, the theorem states that for an infinitely slow
parametric perturbation there is no possibility of a quantum
jump~\cite{born1927dai}. Many physical phenomena observed in both quantum and
classical systems can be explained by this theorem, ranging from optical
tapers~\cite{johnson-joannopoulos2002ata} to robust quantum
gates~\cite{farhi-sipser2001aqa}. Recently, the applicability of adiabatic
principles to non-Hermitian systems, e.g., coupled harmonic modes with gain or
loss, has attracted considerable attention. Here, the complex eigenvalue
structure and the existence of so-called exceptional points
(EPs) leads to new counterintuitive phenomena~\cite{kato1995ptf,
seyranian-mailybaev2003mst, berry2004pon, seyranian-mailybaev2005coe,
rotter2009anh, moiseyev2011nhq, heiss2012tpo, brody-graefe2012mse,
bachelard-sebbah2014coa, peng-yang2014lis, ambichl-rotter2013bop,
klaiman-moiseyev2008vob, ruter-kip2010oop, brandstetter-rotter2014rtp,
dembowski-richter2001eoo}. Perhaps most strikingly, adiabatically encircling
an EP was predicted to effect a state-exchange, with applications for
switching and cooling~\cite{latinne-noble1995ldi, lefebvre-moiseyev2009rci,
atabek-kokoouline2011pfa}. However, it is now known that the very presence of
non-Hermiticity prevents a general application of the adiabatic
theorem~\cite{kvitsinsky-putterman1991aeo, nenciu-rasche1992ota,
dridi-jolicard2010aaf}, and the inevitability of non-adiabatic transitions
leads to new effects, e.g., to chiral behavior~\cite{uzdin-moiseyev2011oto,
berry-uzdin2011snh, gilary-moiseyev2013taq, leclerc-killingbeck2013dot,
graefe-moiseyev2013boa, viennot2014aqa}.

Whereas the above results point to fascinating new physical phenomena, the
complexity of the problem mostly requires one to resort to numerical studies
(as cited above) or to focus on limiting cases where the system evolution is
eventually dominated by a single mode with maximum gain or minimum loss. An
important step beyond this limitation has recently been presented in
Refs.~\cite{berry-uzdin2011snh} and~\cite{berry2011ope} in which an
exactly solvable model is considered and a connection between the appearance
of non-adiabatic transitions and the Stokes phenomenon of
asymptotics~\cite{berry1988uas} is thereby found. However, even for very
simple scenarios, these exact case studies are mathematically already quite
involved, and the translation of the observed dynamics to other systems, in
particular to realistic systems with imperfections and noise, is not
immediately obvious.

In this work we analyse quasi-adiabatic dynamics in non-Hermitian systems near
EPs with the aim to provide a generalised framework for both modelling and
understanding the associated dynamical phenomena. Our approach reveals that
the solutions are in general composed of periods of quasi-stationary during
which the solution follows fixed points, interrupted by abrupt non-adiabatic
transitions due to the exchange of stability. However, the time of these
transitions cannot be predicted by a standard stability analysis, and,
intriguingly, we find that piece-wise adiabaticity is still a key ingredient
for understanding the evolution of the system in spite of an overall breakdown
of adiabatic principles.

On a more fundamental level, our analysis shows that the quasi-adiabatic
dynamics near an EP is a singluarly perturbed
problem~\cite{berglund-gentz2006nip}, meaning that, in contrast to Hermitian
systems, the dynamics cannot be obtained by perturbative corrections to the
adiabatic prediction. This fact makes adiabatic principles in non-Hermitian
systems particularly interesting as well as challenging to understand, both
from a physical and from a mathematical point of view. Specifically, here we
connect the problem of non-adiabatic transitions to the more general
phenomenon of stability loss delay~\cite{diener1984tcu, neishtadt2009osl} in
dynamical bifurcations. This concept more easily affords intuition in
complicated examples where exact solutions cannot be found and in realistic
systems where noise cannot be ignored. Our results are therefore important for a
variety of modern-day experiments with, e.g.,
waveguides~\cite{klaiman-moiseyev2008vob, ruter-kip2010oop}, coupled
resonators~\cite{brandstetter-rotter2014rtp, dembowski-richter2001eoo},
semiconductor microcavities~\cite{gao-ostrovskaya2015oon}, or
electromechanical~\cite{faust-weig2013cco, okamoto-yamaguchi2013cpm} and
optomechanical systems~\cite{xu-li2014mpt, shkarin-harris2014omh,
jing-nori2014pts}, which offer sufficiently high control for the observation
of the predicted dynamical phenomena.

\begin{figure}
	\begin{center}
		\includegraphics[width=\columnwidth]{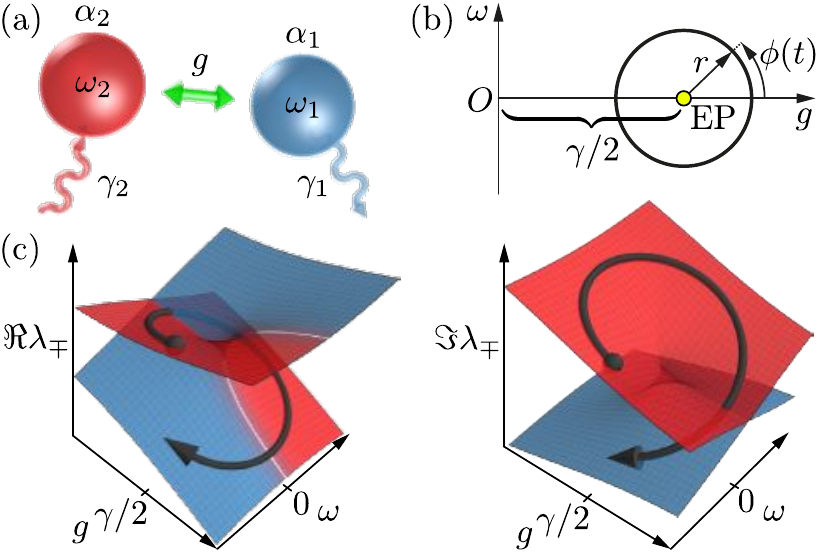}
		\caption{(Color online.) (a)~Cartoon of two coupled harmonic modes
with gain or loss. (b)~Example parametric path where $\gamma$ is fixed,
$\omega = r \sin \phi(t)$, and $g = \gamma / 2 + r \cos \phi(t)$. (c)~Real
($\Re$) and imaginary ($\Im$) parts of the spectrum, $\lambda_\mp = \mp
\sqrt{(\omega + i \gamma / 2)^2 + g^2}$. The curve is the trajectory of
$\lambda_-$ for the path defined in~(b) and depicts the fully adiabatic
evolution.}
		\label{fig:introduction picture}
	\end{center}
\end{figure} 

\begin{figure*}[tbf]
	\begin{center}
		\includegraphics[width=\textwidth]{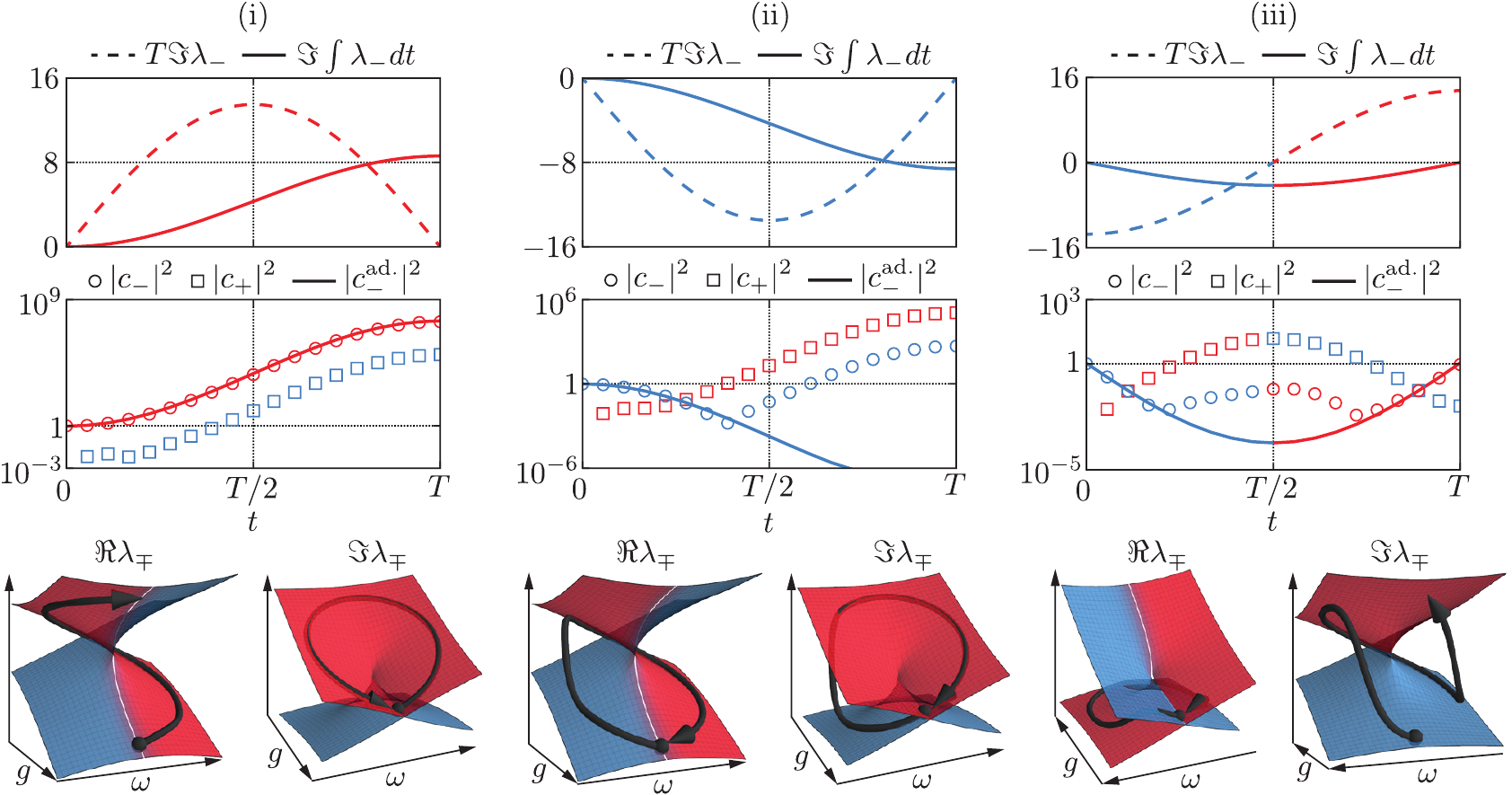}
		\caption{(Color online.) Plots of typical numerical solutions of
Eq.~\eqref{eq:model} for the path defined in
Fig.~\ref{fig:introduction picture}(b) with initial eigenvector populations
$c_-(0) = 1$ and $c_+(0) = 0$. For the function $\phi$ we choose $\phi(t) =
\pm 2 \pi t / T$ in examples~(i,~ii), and we choose $\phi(t) = - 2 \pi t / T +
\pi$ in example~(iii). In all cases we set $r = 0.1$, $\gamma = 1$ and $T =
45$, for which $T |\lambda_- - \lambda_+| \gg 1$. The top row shows the
dynamical gain parameter, $T \Im\lambda_-(t)$, and the total integrated gain,
$\int_0^t \Im\lambda_-(t') dt'$. Note that the dynamical gain is the gain of
the adiabatic prediction but not necessarily the actual gain of the numerical
solution. The middle row shows the eigenvector populations, $|c_\mp(t)|^2$,
along with the adiabatic prediction, $|c_-^{\rm ad.}(t)|^2$. We do not plot
$|c_+^{\rm ad.}(t)|^2$ because adiabatic principles imply $|c_+^{\rm
ad.}(t)|^2 \ll |c_-^{\rm ad.}(t)|^2$. The bottom row shows a projection of the
numerical solution onto the real and imaginary parts of the eigenspectrum,
specifically $[|c_-(t)|^2 \lambda_-(t) - |c_+(t)|^2 \lambda_+(t)] /
[|c_-(t)|^2 + |c_+(t)|^2]$. The use of red and blue is to provide an
indication of which population, or surface, corresponds to a gain and loss
eigenvector respectively.}
		\label{fig:numerical examples}
	\end{center}
\end{figure*}

\section{Non-Hermitian dynamics and exceptional points}
\label{sec:Non-Hermitian dynamics and exceptional points}

\subsection{Model}
\label{sec:Model}

For the following discussion we consider the generic model of two coupled
harmonic oscillators with frequencies $\omega_1$ and $\omega_2$, decay rates
$\gamma_1$ and $\gamma_2$, and coupling strength $g$; see
Fig.~\ref{fig:introduction picture}(a). The equations of motion for the
amplitudes $\alpha_1$ and $\alpha_2$ are
\begin{equation}
	\frac{d}{dt} \begin{pmatrix} \alpha_1 \\ \alpha_2 \end{pmatrix} = - i \begin{pmatrix}
		\omega_1 - i \gamma_1 / 2 & g \\
		g & \omega_2 - i \gamma_2 / 2
	\end{pmatrix} \begin{pmatrix} \alpha_1 \\ \alpha_2 \end{pmatrix},
\end{equation}
where in general $\omega_i = \omega_i(t)$, $\gamma_i = \gamma_i(t)$, and $g =
g(t)$ are functions of time. For the following analysis it is convenient to
eliminate the common evolution with average frequency $\Omega := (\omega_2 +
\omega_1) / 2$ and average decay rate $\Gamma := (\gamma_2 + \gamma_1) / 2$ by
introducing a new set of amplitudes $\beta_1$ and $\beta_2$ via  
\begin{equation}\label{eq:Transformation}
	\begin{pmatrix}
		\alpha_1(t) \\ \alpha_2(t)
	\end{pmatrix} =
		e^{- i \int^t [\Omega(t') - i \Gamma(t') / 2] dt'} \begin{pmatrix}
		\beta_1(t) \\ \beta_2(t)
	\end{pmatrix} .
\end{equation}
The remaining non-trivial dynamics in this frame is 
\begin{gather}\label{eq:model}
	\frac{d}{dt} \begin{pmatrix} \beta_1 \\ \beta_2 \end{pmatrix} = - i \begin{pmatrix}
		- \omega - i \gamma / 2 & g \\
		g & \omega + i \gamma / 2
	\end{pmatrix} \begin{pmatrix} \beta_1 \\ \beta_2 \end{pmatrix},
\end{gather}
where $\omega := (\omega_2 - \omega_1) / 2$ and $\gamma := (\gamma_1 -
\gamma_2) / 2$. Note that while the global
transformation~\eqref{eq:Transformation} does not affect any of the following
results, if $\Gamma \neq 0$ then the experimentally observable amplitudes
$\alpha_{1, 2}$ are related to $\beta_{1, 2}$ by an exponentially large or
small prefactor.

Below we suppose that at least $\omega$ and $g$, or $\omega$ and $\gamma$ can
be controlled as a function of time. This can be achieved, e.g., with optical
modes propagating through waveguides with spatially varying
losses~\cite{uzdin-moiseyev2012sfa, doppler-rotter2014}, by applying chirped
laser pulses to molecular systems~\cite{gilary-moiseyev2013taq}, or by using
two mechanical resonators with electrically~\cite{faust-weig2013cco,
okamoto-yamaguchi2013cpm} or optomechanically~\cite{xu-li2014mpt,
shkarin-harris2014omh, jing-nori2014pts} controlled parameters.

\subsection{Exceptional points}
\label{sec:Exceptional points}

Let us write Eq.~\eqref{eq:model} more compactly as $\dot{\vec{x}} = - i
\textbf{M} \vec{x}$, where $\vec{x}$ is the state vector and $\textbf{M}$ is
the dynamical matrix, or sometimes called in this context a non-Hermitian
Hamiltonian~\cite{moiseyev2011nhq}, i.e.,
\begin{equation}
	\vec{x} := \begin{pmatrix}
		\alpha_1 \\ \alpha_2
	\end{pmatrix} \text{ and } {\bf M} := \begin{pmatrix}
		- \omega - i \gamma / 2 & g \\
		g & \omega + i \gamma / 2
	\end{pmatrix} .
\end{equation}
${\bf M}$ has eigenvalues $\lambda_\mp = \mp \lambda = \mp \sqrt{(\omega + i
\gamma / 2)^2 + g^2}$. The corresponding eigenvectors are
\begin{gather}\label{eq:eigenvectors}
	\vec{r}_- = \begin{pmatrix} \cos \vartheta / 2 \\ \sin \vartheta / 2 \end{pmatrix}
		\text{ and }
		\vec{r}_+ = \begin{pmatrix} - \sin \vartheta / 2 \\ \cos \vartheta / 2 \end{pmatrix}
\end{gather}
with $\vartheta$ such that $\tan \vartheta = - g / (\omega + i \gamma /
2)$. Figure~\ref{fig:introduction picture}(c) shows the real
($\Re$) and imaginary ($\Im$) parts of $\lambda_\pm$ as a function of $g$ and
$\omega$ with $\gamma$ fixed. The pinch points $\omega + i \gamma / 2 \mp i g
= 0$ are EPs~\cite{kato1995ptf, seyranian-mailybaev2003mst,
seyranian-mailybaev2005coe, moiseyev2011nhq, berry2004pon, rotter2009anh,
heiss2012tpo}. At these points the eigenvalues as well as the eigenvectors
coalesce, and ${\bf M}$ becomes non-diagonalizable. Encircling an EP with a
closed path in parameter space causes the two eigenvalues, and hence also the
two eigenvectors, to swap; see Figs.~\ref{fig:introduction picture}(b,~c).
Based on intuition from the quantum adiabatic theorem, it was suggested that
this unique feature could be observed in physical systems by encircling an EP
over a time $T$ such that $T |\lambda_- - \lambda_+|$ is
large~\cite{latinne-noble1995ldi, lefebvre-moiseyev2009rci,
atabek-kokoouline2011pfa}. However, other studies contradict this result and
show that due to non-Hermiticity this picture cannot hold in
general~\cite{kvitsinsky-putterman1991aeo, nenciu-rasche1992ota,
dridi-jolicard2010aaf, uzdin-moiseyev2011oto, berry-uzdin2011snh,
gilary-moiseyev2013taq, leclerc-killingbeck2013dot, graefe-moiseyev2013boa,
viennot2014aqa}.

\subsection{Numerical examples}
\label{sec:Numerical examples}

Before presenting a further analytic treatment of Eq.~\eqref{eq:model} we
consider in Fig.~\ref{fig:numerical examples} some typical solutions for
encircling an EP with $T |\lambda_- - \lambda_+| \gg 1$. For these examples we
choose a path in parameter space as defined in
Fig.~\ref{fig:introduction picture}(b). We expand the solution as
\begin{equation}\label{eq:EVDecomposition}
	\vec{x}(t) = c_-(t) \vec{r}_-(t) + c_+(t) \vec{r}_+(t) ,
\end{equation}
where $\vec{r}_-(t)$ and $\vec{r}_+(t)$ are the \emph{instantaneous}
eigenvectors of ${\bf M}(t)$, and we choose the initial condition $c_-(0) = 1$
and $c_+(0) = 0$. The adiabatic prediction is $c_-^\text{ad.}(t) \simeq e^{- i
\int_0^t \lambda_-(t') dt'}$ and $c_+^\text{ad.}(t) \ll
c_-^\text{ad.}(t)$~\cite{fn:geometric_phase, leclerc-jolicard2012tro,
ibenez-muga2014acf}.

In examples~(i) and~(ii) in Fig.~\ref{fig:numerical examples} we have chosen
an anticlockwise and a clockwise encircling respectively, $\phi(t) = \pm 2 \pi
t / T$. In the anticlockwise example the solution matches the adiabatic
prediction and the corresponding state flips, but in the clockwise example we
observe a non-adiabatic transition, for which, apart from an overall
amplification, the system returns to the original state. This chiral behavior,
first presented in Ref.~\cite{uzdin-moiseyev2011oto}, illustrates one of the
key differences between the dynamics in Hermitian and non-Hermitian systems.
In the latter, the eigenvalues are complex, which causes gain or loss in $c_-$
and $c_+$. An infinitesimally small non-adiabatic coupling can therefore be
exponentially amplified, causing the gain eigenvector to dominate. This
mechanism intuitively explains why the adiabatic theorem does not in general
hold for non-Hermitian systems.

Example~(iii) shows the result for a more interesting path $\phi(t) = - 2 \pi
t / T + \pi$ where gain-loss behavior swaps half-way through and the total
integrated dynamical gain vanishes, $\int_0^T \Im \lambda(t) dt = 0$.
Surprisingly, the final state matches the adiabatic prediction, $|c_-(T)|^2
\simeq |c_-(0)|^2$, even though during the interim the solution is highly
non-adiabatic. This observation cannot be explained by the intuitive argument
above because $c_-$ is non-trivially slaved to $c_+$ past the time $t = T / 2$
when we would expect $c_-$ to increase exponentially. Thus, considering
dynamical gain alone is insufficient to accurately predict behaviour for
quasi-adiabatic dynamics near EPs.

These basic examples illustrate that the dynamics of non-Hermitian systems
involves three characteristic effects: 
\begin{enumerate}[(i)]
	\item The swapping of eigenvectors due to a $4 \pi$-periodicity about an
EP, which follows from the topology of the complex eigenvalue spectrum.
	\item The appearance of enhanced non-adiabatic transitions due to the
presence of gain or loss. 
	\item Periods of adiabatic evolution that persist significantly beyond the
time of stability loss.
\end{enumerate} 
While~(i) is readily incorporated by the eigenvector
decomposition~\eqref{eq:EVDecomposition}, we will now develop a general
approach to describe the non-trivial interplay between~(ii) and~(iii).

\section{Dynamical analysis}
\label{sec:Dynamical analysis}

\subsection{Relative non-adiabatic transition amplitudes}
\label{sec:Relative non-adiabatic transition amplitudes}

In order to develop a general dynamical description we consider the evolution
operator, $\mathcal{U}(t)$, defined by $\vec{x}(t) = \mathcal{U}(t)
\vec{x}(0)$, which contains the full dynamics independent of the initial
condition. In the eigenbasis Eq.~\eqref{eq:eigenvectors}, $\mathcal{U}(t)$ is
the solution of
\begin{gather}\label{eq:evolution operator}
	\dot{\mathcal{U}} = - i \begin{pmatrix}
		- \lambda(t) & - f(t) \\
		f(t) & \lambda(t)
	\end{pmatrix} \mathcal{U} \text{,}\quad \mathcal{U} = \begin{pmatrix}
		\mathcal{U}_{-, -} & \mathcal{U}_{-, +} \\
		\mathcal{U}_{+, -} & \mathcal{U}_{+, +}
	\end{pmatrix} 
\end{gather}
with initial condition $\mathcal{U}(0) = \mathbbm{1}$, where
\begin{gather}
	f(t) = \frac{
			g(t) \big[ \dot{\omega}(t) + i \dot{\gamma}(t) / 2 \big]
			- \big[ \omega(t) + i \gamma(t) / 2 \big] \dot{g}(t)
		}{
			2 i \lambda^2(t)
		} ,
\end{gather} 
is the non-adiabatic coupling~\cite{fn:geometric_phase}. Adiabaticity
usually requires that the non-adiabatic coupling be much smaller than the
distance between eigenvectors, $\varepsilon(t) := |f(t) / [2
\lambda(t)]| \ll 1$. Since $\varepsilon(t) \propto T^{-1}$ this condition is
always satisfied for an appropriate $T$. To set $f(t) = 0$ in
Eq.~\eqref{eq:evolution operator}, which would imply $\varepsilon(t) = 0$,
would yield the diagonal adiabatic prediction
\begin{equation}
	\mathcal{U}^{\text{ad.}}(t) = \begin{pmatrix}
		e^{- i \int_0^t \lambda(t') dt'} & 0 \\
		0 & e^{i \int_0^t \lambda(t') dt'}
	\end{pmatrix}.
\end{equation}
However, as is evident in Fig.~\ref{fig:numerical examples}, even for
arbitrarily small yet non-vanishing $\varepsilon(t)$ the actual solution is
significantly non-diagonal. This indicates that the system is \emph{singularly
perturbed} by the non-adiabatic coupling, and $\mathcal{U}(t)$ cannot be
obtained as a perturbative correction to $\mathcal{U}^{\text{ad.}}(t)$. We
shall henceforth call $\varepsilon(t) \ll 1$ the \emph{quasi}-adiabatic
condition (see Appendix~\ref{sec:Quasi-adiabaticity} for more details).

In order to describe the non-adiabatic character of $\mathcal{U}(t)$ for
quasi-adiabatic dynamics we focus on the relative non-adiabatic transition
amplitudes~\cite{uzdin-moiseyev2011oto}:
\begin{gather}\label{eq:definition of R}
	R_-(t) := \frac{\mathcal{U}_{+, -}(t)}{\mathcal{U}_{-, -}(t)}
		\text{ and }
	R_+(t) := \frac{\mathcal{U}_{-, +}(t)}{\mathcal{U}_{+, +}(t)} .
\end{gather}
These describe the amount of non-adiabaticity in the solution. For example,
$R_-(t)$ is a measure of the magnitude of the net non-adiabatic transition
from $\vec{r}_-(t)$ to $\vec{r}_+(t)$. If $R_\mp(t) \ll 1$ then we may say
that $c_\mp$ is behaving adiabatically, while $R_\mp(t) \gg 1$ indicates that
a non-adiabatic transition has occurred. From
Eqs.~\eqref{eq:evolution operator} and~\eqref{eq:definition of R} it follows
that $R_\mp(t)$ considered as a dynamical variable is the solution to the
Riccati equation~\cite{gradshteyn-ryzhik2007toi, berry2011ope}
\begin{gather}\label{eq:dynamical flip errors}
	\dot{R}_\mp = \mp 2 i \lambda(t) R_\mp \mp i f(t) (1 + R_\mp^2)
\end{gather}
with initial condition $R_\mp(0) = 0$. Dynamical phenomena associated with
quasi-adiabatically encircling EPs can thus be understood from the solutions
of this equation in the limit $\varepsilon(t) \ll 1$. Note that the equations
of motion for $R_-$ and $R_+$ are related via $R_- \leftrightarrow 1 / R_+$.
In the following, we therefore consider only $R := R_-$ without loss of
generality.

We remark that, assuming transients are damped, the relation $R_-
\leftrightarrow R_+$ has the immediate consequence that $\lim_{t \rightarrow
\infty} R_-(t) R_+(t) = 1$, which agrees with
Ref.~\cite{uzdin-moiseyev2011oto} and prohibits simultaneous adiabatic
behaviour in both $c_-$ and $c_+$ over long times. 

\subsection{Fixed points and stability loss delay}
\label{sec:Fixed points and stability loss delay}

\begin{figure}[tbf]
	\begin{center}
		\includegraphics[width=\columnwidth]{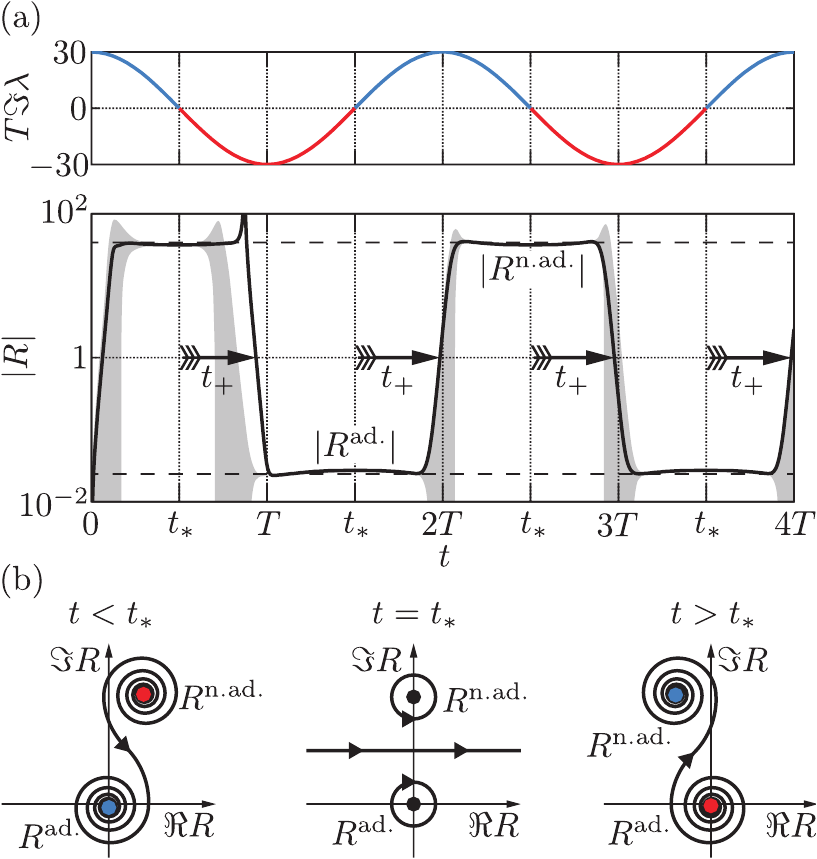}
		\caption{(Color online.) (a)~Plot of $\Im \lambda(t)$ (upper panel),
and a typical solution for $R \equiv R_-$ (lower panel), for the path defined
in Fig.~\ref{fig:introduction picture}(b) with $\phi(t) = - 2 \pi t / T +
\pi$. Note that $\Im \lambda(t) = - \Im \lambda_-(t)$, which is plotted in
Fig.~\ref{fig:numerical examples}. We have chosen $r = 0.1$, $\gamma = 1$ and
$T = 100$, for which $\varepsilon(t) \simeq 2.5\%$. The solid curve is the
numerical solution. The arrows denote delay times. The lower and upper dashed
grid lines denote $|R^\text{ad.}(t)|$ and $|R^\text{n.ad.}(t)|$ respectively.
The shaded area is one standard deviation about the mean of $R_-$ obtained
from $10{,}000$ stochastic numerical integrations of $c_-$ and $c_+$ (see
Sec.~\ref{sec:Noise} for more details). (b)~Cartoons of the global phase
portraits of the equation of motion for $R$ near $t_*$. Arrows denote the
direction of time-evolution along an integral curve. The fixed point near the
origin corresponds to $R^\text{ad.}(t)$, and the fixed point far from the
origin corresponds to $R^\text{n.ad.}(t)$.}
		\label{fig:figure3}
	\end{center}
\end{figure}

The lower panel of Fig.~\ref{fig:figure3}(a) shows a generic solution for $R$
during multiple quasi-adiabatic encirclements of an EP (see the caption for
details). It resembles a square wave, i.e., we see fast switching between two
quasi-stationary values. This behavior can be understood from a separation of
time-scales in Eq.~\eqref{eq:dynamical flip errors}. For short times the
slowly varying parameters $\lambda(t) \simeq \lambda$ and $f(t)\simeq f$ can
be considered constant and 
\begin{equation}
	\dot{R} \simeq - 2 i \lambda R - i f (1 + R^2) .
\end{equation}
On a fast time-scale set by $|\Im \lambda|^{-1}$ the solution therefore
approaches one of two fixed points 
\begin{gather}
	\begin{aligned}
		R^{\text{ad.}} &= - \frac{\lambda}{f} \left(
			1 - \frac{\sqrt{\lambda^2 - f^2}}{\lambda}
			\right) \simeq - \frac{f}{2 \lambda} \text{ and}\\
		R^{\text{n.ad.}} &= - \frac{\lambda}{f} \left(
			1 + \frac{\sqrt{\lambda^2 - f^2}}{\lambda}
			\right) \simeq - \frac{2 \lambda}{f} .
	\end{aligned}
\end{gather}
The first fixed point, $R^{\rm ad.}(t) \propto \varepsilon(t) \ll 1$,
indicates adiabatic behavior ($c_-$ dominates) and is stable for
$\Im\lambda(t) < 0$. The second fixed point, $R^{\rm n.ad.}(t) \propto
\varepsilon^{-1}(t) \gg 1$, indicates a non-adiabatic transition has occured
($c_+$ dominates) and is stable for $\Im \lambda(t) > 0$. These two fixed
points are plotted in Fig.~\ref{fig:figure3}(a). Evidently, the periods of
quasi-stationarity there exhibited correspond to following one of these two
fixed points.

On a slow time-scale set by $T$ the parameters $\lambda(t)$ and $f(t)$ may
change considerably and at certain critical times the stability of the two
fixed points swaps. For example, $R^{\text{ad.}}(t)$ becomes unstable and
$R^{\text{n.ad.}}(t)$ stable when the sign of $\Im \lambda(t)$ becomes
positive. Let us denote the critical times by $t_*$, which are marked in
Fig.~\ref{fig:figure3}(a). Na{\"i}vely, one might expect an immediate rapid
transition between the neighbourhoods of $R^{\text{ad.}}(t)$ and
$R^{\text{n.ad.}}(t)$ upon passing a critical time $t_*$, but, as is evident
in Fig.~\ref{fig:figure3}(a), this is not the case. Instead, we see that the
solution follows, e.g., $R^{\text{ad.}}(t)$ while it is unstable for a
significant amount of time; the loss in stability is delayed. Intuition for
this behaviour is obtained from the phase portraits of
Eq.~\eqref{eq:dynamical flip errors}, shown in Fig.~\ref{fig:figure3}(b). The
local phase portrait about $R^{\text{ad.}}(t)$ goes from a spiral towards
$R^{\text{ad.}}(t)$ for $t < t_*$ to a spiral away from $R^{\text{ad.}}(t)$
for $t > t_*$, passing through a degenerate bifurcation at $t = t_*$ when
$R^{\text{ad.}}(t)$ is a centre and neither stable nor unstable. We therefore
expect some persistence in the following of $R^{\text{ad.}}(t)$ because near
$t_*$ it is only `weakly' stable or unstable.

To illustrate the existence of a significant delay between the critical time
$t_*$ and the actual time of a non-adiabatic transition $t_+$ we consider the
specific path defined in Fig.~\ref{fig:introduction picture}(b) with $\phi(t)
= - 2 \pi t / T + \pi$ and $r \ll \gamma$. This is a good model for the
numerical solution shown in Fig.~\ref{fig:figure3}(a). Then, $\lambda(t)
\simeq i \sqrt{r \gamma} e^{- i \pi t / T}$, $f(t) \simeq i \pi / (2 \, T)$,
and $\varepsilon(t) \simeq \varepsilon = \pi / (4 \sqrt{r \gamma} \, T)$. Let
us focus on the loss of stability of $R^{\text{ad.}}(t)$ at $t_* = 3 T / 2$.
Assuming that the system is near $R^{\rm ad.}(t)$ we can neglect the nonlinear
term in Eq.~\eqref{eq:dynamical flip errors}:
\begin{equation}
	\frac{\dot{R}}{2 \sqrt{r \gamma}} \simeq e^{- i \pi t / T} R + \varepsilon .
\end{equation}
The particular integral of this equation is found to be
\begin{equation}
	R(t) = - \frac{i}{2} E_1\left( \frac{i}{2 \varepsilon} e^{- i \pi t / T} \right)
		e^{i \exp(- i \pi t / T) / (2 \varepsilon)} ,
\end{equation}
where $E_1$ is the exponential integral. Since $\varepsilon \ll 1$ we may use
the asymptotic expansion for $E_1$ (see, e.g., 5.1.7 and 5.1.51 in
Ref.~\cite{abramowitz-stegun1972hom}) to obtain
\begin{gather}\label{eq:asymptotic solution}
	\begin{aligned}
		R(t) &\simeq - \varepsilon e^{i \pi t / T}
			- 2 i \varepsilon^2 e^{2 i \pi t / T} + \dots \\
		&- \frac{\pi}{2} \operatorname{sgn}\left[ \cos\left( \frac{\pi t}{T} \right) \right]
			e^{i \exp(- i \pi t / T) / (2 \varepsilon)} .
	\end{aligned}
\end{gather}%
The first two terms (upper line on the right) correspond to following $R^{\rm
ad.}(t)$ with higher order corrections. The third term (lower line) is
negligible for $(t - t_*) < T / 2$ (recall $t_* = 3 T / 2$ here), but it
diverges exponentially for $(t - t_*) > T / 2$, thereby indicating a
non-adiabatic transition. Thus, under the ideal conditions assumed here and
given that the solution has approached $R^{\text{ad.}}(t)$ by $t = t_*$, the
delay in the loss of stability is $(t_+ - t_*) = T / 2$.

With this analysis we are already in a position to understand better the three
examples studied in Fig.~\ref{fig:numerical examples}. In example~(i)
$R^{\text{ad.}}(t)$ is stable for the entire loop around the EP and therefore
the solution follows the adiabatic prediction, $|c_+(t) / c_-(t)| \simeq
|R^{\text{ad.}}(t)|$. In contrast, in~(ii) $R^{\text{ad.}}(t)$ is always
unstable and a non-adiabatic transition occurs. In~(iii) the
solution first switches from $R^{\text{ad.}}(t)$ to $R^{\text{n.ad.}}(t)$, but
then back again with a delay $t_+ \lesssim T / 2$ after $R^{\text{ad.}}(t)$
becomes stable at $t = T / 2$. Note that the delay times exhibited in the
first encircling period as seen in Figs.~\ref{fig:numerical examples}
and~\ref{fig:figure3}(a) differ somewhat from the value $t_+ = T / 2$
estimated above. This is due to a high sensitivity to the initial condition
$R(0) = 0$, which is not exponentially close to $R^{\rm ad.}(0)$, and
therefore effects a transient term of the form $A e^{i \exp(- i \pi t / T) /
(2 \varepsilon)}$. After about one encircling period the system approaches the
unique long-time relaxation oscillation which is a universal signature of
quasi-adiabatically encircling EPs.

We finish this section with a remark on the relation between the above results
and the \emph{Stokes phenomenon of asymptotics}, i.e., the switching-on of
exponentially suppressed terms in asymptotic expansions~\cite{berry1988uas}.
In Refs.~\cite{berry-uzdin2011snh, berry2011ope} an exact solution for the
example considered in this subsection is presented (using $r \ll \gamma$ but
not neglecting the nonlinearity), which we review in
Appendix~\ref{sec:Non-adiabatic transitions as a manifestation of the Stokes phenomenon of asymptotics}.
In this exact solution one sees that the sharp (but continuous) transition,
which in Eq.~\eqref{eq:asymptotic solution} is represented by the signum
function, is precisely the Stokes phenomenon of asymptotics, leading here to
a breakdown of the adiabatic theorem. In our current approach, which we
elaborate further in the next section, this discontinuity is connected to the
problem of stability loss delay. To our knowledge the connection between the
Stokes phenomenon of asymptotics and stability loss delay has hitherto not
been suggested, and might be worth exploring further. However, here we will
leave such considerations aside and proceed with a pragmatic generalisation of
these initial results to arbitrary paths in parameter space. 

\subsection{Generalised quasi-adiabatic solution}
\label{sec:Generalised quasi-adiabatic solution}

In the previous section, Sec.~\ref{sec:Fixed points and stability loss delay},
we were able to understand the generic solution exhibited in
Fig.~\ref{fig:figure3}(a) from a separation of time-scales, which resulted in
a delay in the loss of stability of the instantaneous fixed points. In fact,
slow-fast systems with dynamical bifurcations are a subject of current
mathematical interest. The reader is referred to
Ref.~\cite{berglund-gentz2006nip} for a concise description. The reason that
the critical times do not coincide with the observed times when an
instantaneous fixed point loses stability is because our slow-fast system is
\emph{singularly perturbed}; the slow system is described by an algebraic
equation, while the fast system by a differential equation. One must therefore
resort to non-standard analysis. A principal result of the non-standard
analysis of slow-fast systems is the existence of \emph{stability loss delay}
about certain dynamical bifurcations~\cite{diener1984tcu, neishtadt2009osl,
bohun2011sld}, which we observed explicitly in
Sec.~\ref{sec:Fixed points and stability loss delay}. In the following we
build upon this to construct a generalised quasi-adiabatic solution, which,
additionally, affords an estimation of delay times.

We are interested in solutions that for times near critical times $t_*$ are in
the vicinity of a fixed point. We therefore begin by looking at the
zero-crossings of $\Im \lambda(t)$, which determine $t_*$. For some window
$[t_-, t_+]$ about each $t_*$, i.e., $t_- < t_* < t_+$, we seek a solution
$R_{t_*}(t)$ that follows $R^{\text{ad.}}(t)$ or $R^{\text{n.ad.}}(t)$. Since
transitions between $R^{\text{ad.}}(t)$ and $R^{\text{n.ad.}}(t)$ are very
quick, by making a piece-wise addition of segments that follow one or the
other fixed point we arrive at the approximation for the complete solution
thus
\begin{equation}\label{eq:Stiching}
	R(t) \simeq \sum_{t_*} [\Theta(t - t_-) - \Theta(t - t_+)] R_{t_*}(t),
\end{equation}
where $\Theta$ is the Heaviside step function.

Let us now consider a single segment and omit the subscript $t_*$ for brevity.
We may focus on the case that $R(t)$ follows $R^{\text{ad.}}(t)$ without loss
of generality because $R^{\text{n.ad.}}(t) = 1 / R^{\text{ad.}}(t)$ and
Eq.~\eqref{eq:dynamical flip errors} is antisymmetric under the transformation
$R \mapsto 1 / R$. Since we assume $R(t)$ to be in the vicinity of
$R^{\text{ad.}}(t)$ for $t \in [t_-, t_+]$ we study the linearised equation of
motion about $R^{\text{ad.}}(t)$:
\begin{equation}
	\dot{R} = - 2 i \lambda(t) R - i f(t) .
\end{equation}
The general solution from time $t = t_0$ of this equation is
\begin{equation}\label{eq:general solution}
	R(t) = R(t_0) e^{\Psi(t) - \Psi(t_0)} - i \int_{t_0}^{t} dt' f(t') e^{\Psi(t) - \Psi(t')},
\end{equation}
where $R(t_0)$ is the initial condition and
\begin{equation}
	\Psi(t) = - 2 i \int_{t_*}^{t} \lambda(t') dt' .
\end{equation}
Note that, to first order about $t_*$ we have $\lambda(t) = \lambda(t_*) +
\dot{\lambda}(t_*) (t - t_*) + \mathcal{O}((t - t_*)^2)$. Since $\Im
\lambda(t_*) = 0$ and $\Im \dot{\lambda}(t_*) > 0$ then $\Re \Psi(t) = \Im
\dot{\lambda}(t_*) (t - t_*)^2 + \mathcal{O}((t - t_*)^3)$ is convex. We refer
to this property of $\Psi$ below. Integrating the integral in
Eq.~\eqref{eq:general solution} by parts $N$~times yields
\begin{equation}\label{eq:Rsolution}
	\begin{aligned}
		R(t) =& [R(t_0) - \mathcal{R}^{\text{ad.}}(t_0)] e^{\Psi(t) - \Psi(t_0)} \\
		&+ \mathcal{R}^{\text{ad.}}(t) + \Delta(t) e^{\Psi(t)}.
	\end{aligned}
\end{equation}
Here we have introduced
\begin{equation}\label{eq:SumRad}
	\mathcal{R}^{\text{ad.}}(t) = \sum_{n = 0}^{N - 1} \left(
		\frac{-1}{2 i \lambda(t)} \frac{d}{dt}
		\right)^n R^{\text{ad.}}(t),
\end{equation}
which encapsulates the following of $R^{\text{ad.}}(t)$: The $n = 0$ term in
$\mathcal{R}^{\text{ad.}}(t)$ is simply $R^{\text{ad.}}(t)$, and the higher
order terms are corrections due to finite variations in $\lambda(t)$ and
$f(t)$. However, since each term in the sum contains a derivative and
therefore scales with $n!$, there is an optimal truncation $N =
N_{\text{op.}}$ beyond which the sum diverges. The precise value of
$N_{\text{op.}}$ is problem-specific, but for most purposes including only the
first few terms in the sum Eq.~\eqref{eq:SumRad} is sufficient.

The final term in Eq.~\eqref{eq:Rsolution}, $\Delta(t) e^{\Psi(t)}$, is the
remaining part of the solution which is not included in the sum
Eq.~\eqref{eq:SumRad}. It therefore describes the non-trivial part of
the dynamics that inevitably causes a departure from $R^{\text{ad.}}(t)$.
Since $\Delta(t) e^{\Psi(t)}$ is the remainder of an optimally truncated sum
it is negligible whenever the solution follows $R^{\text{ad.}}(t)$. On the
other hand, for times $t \approx t_+$ when $\Delta(t) e^{\Psi(t)}$ starts to
dominate, $\mathcal{R}^{\text{ad.}}(t)$ is negligible and we may approximate
\begin{equation}\label{eq:Deltat}
	\Delta(t) e^{\Psi(t)} \simeq - i  e^{\Psi(t)} \int_{t_0}^{t} dt' f(t') e^{- \Psi(t')} .
\end{equation}
Since $\Re \Psi$ is convex and since $\Psi(t) \propto \varepsilon^{-1}(t)$,
the integrand in Eq.~\eqref{eq:Deltat} is non-negligible only for times $t'
\approx t_*$ and the value of $\Delta$ becomes quite independent of $t > t_*$
and $t_0 < t_*$. Therefore, under quite general conditions, we can approximate
$\Delta(t) e^{\Psi(t)} \simeq \Theta(t - t^*) \Delta e^{\Psi(t)}$, where
$\Theta$ is the Heaviside step function, and
\begin{equation}
	\Delta = - i \int_{t_-}^{t_+} dt f(t) e^{- \Psi(t)} .
\end{equation}
The precise values of $t_-$ and $t_+$ are of little importance in the
evaluation of this integral, only that they are far enough from $t_*$ that the
integrand is negligible at them. We thus arrive at
\begin{equation}\label{eq:shadow}
	R(t) \simeq \mathcal{R}^{\text{ad.}}(t) + \left[ A + \Theta(t - t_*) \Delta \right] e^{\Psi(t)} ,
\end{equation}
where $A = [R(t_0) - \mathcal{R}^{\text{ad.}}(t_0)] e^{- \Psi(t_0)}$ depends
on the initial condition.

From Eq.~\eqref{eq:shadow}, and the analogous expression for a segment that
follows $R^{\text{n.ad.}}(t)$, we construct our piece-wise addition of
segments by determining the exit time $t_+$ of a segment from the condition
$|R(t_+)| = 1$, i.e. when the solution is `half-way between'
$R^{\text{ad.}}(t)$ and $R^{\text{n.ad.}}(t)$, and then using this as the
entry time $t_-$ for the next segment. Two effects may cause this transition.
Firstly, if the solution does not have enough time to approach, e.g.,
$R^{\text{ad.}}(t)$ sufficiently closely by the critical time, then the finite
difference $|R(t_*) - R^{\text{ad.}}(t_*)|$ will be exponentially amplified
after $t = t_*$. This mechanism is responsible, e.g., for the initial
transitions one observes in a single encircling of an EP, where the system is
initialised to $R(0) = 0 \not\approx R^{\text{ad.}}(t)$. Secondly, however, we
see that even for $A = 0$ a destabilisation occurs due to a dynamical
mechanism represented by $\Delta \neq 0$, which yields the time of stability
loss $t_+$ via
\begin{equation}\label{eq:departure time}
	|\Delta e^{\Psi(t_+)}| = 1 .
\end{equation}
The time $t_+$ determined in this way is independent of transients and
therefore characterises the longest time the solution can remain stable after
$t_*$. In the quasi-adiabatic limit $\varepsilon(t) \rightarrow 0$ this is not
only independent of transients but also of adiabaticity, and is in fact the
so-called \emph{maximal delay time} $t_+^*$ (see
Appendix~\ref{sec:Stability loss delay} for more details).

\subsection{Analytic examples}
\label{sec:Analytic examples}

Here we consider three examples analytically in order to illustrate our
generalised quasi-adiabatic solution: a circular $\lambda(t)$ as in
Sec.~\ref{sec:Fixed points and stability loss delay}; a linear $\lambda(t)$
corresponding to the lowest order Taylor expansion; an elliptical $\lambda(t)$
corresponding to the lowest order Fourier expansion. The first example serves
to verify that our generalised quasi-adiabatic solution recovers the more
specific analytic results in
Sec.~\ref{sec:Fixed points and stability loss delay}. The second and third
examples serve to illustrate the sensitivity of $\Delta$ and therefore $t_+$
in Eq.~\eqref{eq:departure time} to the global path---stability loss delay is
a global phenomenon. Note that a circular, elliptical, or linear
$\lambda(t)$ does not precisely correspond to a circular, elliptical, or
linear path in parameter space unless we are in, say, the limit $r \ll 1$. We
study particular paths in parameter space numerically in
Sec.~\ref{sec:Numerical examples 2}.

\emph{Circular $\lambda(t)$.}---From
Sec.~\ref{sec:Fixed points and stability loss delay},
\begin{align}
	\lambda(t) &= i \sqrt{r \gamma} e^{- i \pi t / T} \text{ and}\\
	f(t) &= \frac{i \pi}{2 T} .
\end{align}
The adiabatic fixed point with corrections is
\begin{equation}
	\mathcal{R}^{\text{ad.}}(t)
		= - \varepsilon e^{i \pi t / T} \sum_{n = 0}^{N - 1} n! \left(2 i \varepsilon e^{i \pi t / T}\right)^n
\end{equation}
where $\varepsilon = \pi / 4 \sqrt{r \gamma} T$ and the optimal truncation is
$N \sim (2 \varepsilon)^{-1}$. Furthermore, about $t_* = 3 T / 2$ 
\begin{align}
	\Psi(t) &= \frac{1}{2 \varepsilon} (i e^{- i \pi t / T} + 1) \text{ and}\\
	\Delta &= - \pi e^{- \frac{1}{2 \varepsilon}} .\label{eq:circular discontinuity}
\end{align}
Putting these expressions together in Eq.~\eqref{eq:shadow} recovers
Eq.~\eqref{eq:asymptotic solution}, and solving for $t_+$ in
Eq.~\eqref{eq:departure time} yields the delay time
\begin{equation}
	t_+ - t_* = \frac{T}{\pi} \arccos (2 \varepsilon \log \pi) ,
\end{equation}
which in the limit $\varepsilon \rightarrow 0$ becomes $t_+ - t_* = T / 2$, in
agreement with Sec.~\ref{sec:Fixed points and stability loss delay}.

\emph{Linear $\lambda(t)$.}---Let us now consider another important scenario,
where the line of instability is crossed in a linear sweep,
\begin{align}
	\lambda(t) &= \lambda_{\Re} + i \dot{\lambda}_{\Im} t \text{ and}\\
	f(t) &\simeq f(t_*) = \text{const}
\end{align}
where $\dot{\lambda}_{\Im} > 0$. In this case we have $\Psi(t) = - 2 i
\lambda_{\Re} t + \dot{\lambda}_{\Im} t^2$ about $t_* = 0$ and the
discontinuity is
\begin{equation}\label{eq:linear discontinuity}
	\Delta = - i f(t_*) \sqrt{\frac{\pi}{\dot{\lambda}_{\Im}}} e^{- \lambda_{\Re}^2 / \dot{\lambda}_{\Im}} .
\end{equation}
From these expressions we deduce the delay time
\begin{equation}\label{eq:tp_Linear}
	t_+ - t_* = \frac{\lambda_{\Re}}{\dot{\lambda_{\Im}}} \sqrt{
		1 + \frac{\dot{\lambda}_{\Im}}{\lambda^2_{\Re}} \log \left(
		\sqrt{\frac{\dot{\lambda}_{\Im}}{\pi}} \frac{1}{|f(t_*)|}
		\right)
		} ,
\end{equation}
which, in the quasi-adiabatic limit becomes $t_+ - t_* = \lambda_{\Re} /
\dot{\lambda}_{\Im}$. One might na{\"i}vely hypothesize that
Eq.~\eqref{eq:linear discontinuity} describe more general paths by using
$\lambda_{\Re} = \Re \lambda(t_*)$ and $\lambda_{\Im} = \Im
\dot{\lambda}(t_*)$. However, a comparison with the circular path above
already shows that this would only give rather poor quantitative results.
Eq.~\eqref{eq:tp_Linear} may still serve for a first estimate of the expected
delay times in general scenarios.

%
\emph{Elliptical $\lambda(t)$.}---As an interpolation between the two cases
above we consider the lowest order Fourier expansion of $\lambda(t)$ about $t
= t_*$:
\begin{align}
	\lambda(t) &= \lambda_{\Re} \cos(\pi t / T) + i \frac{T \dot{\lambda}_{\Im}}{\pi} \sin(\pi t / T) \text{ and}\\
	f(t) &\simeq f(t_*) = \text{const} .
\end{align}
With $\lambda_{\Re} = T \dot{\lambda}_{\Im} / \pi = \sqrt{r \gamma}$ one
recovers the circular $\lambda(t)$, and for $T \rightarrow \infty$, but
keeping $\dot{\lambda}_{\Im}$ fixed one recovers the linear sweep of
$\lambda(t)$. For this example the discontinuity is
\begin{equation}\label{eq:Fourier discontinuity}
	\Delta = - 2 i T f(t_*) e^{- \frac{2 T^2 \dot{\lambda}_{\Im}}{ \pi^2}}
		I_0 \left( \frac{2 T}{\pi} \sqrt{\frac{T^2 \dot{\lambda}_{\Im}^2}{\pi^2} - \lambda_{\Re}^2} \right) ,
\end{equation}
where $I_0$ is the first order modified Bessel function of the first kind. By
taking the appropriate limits---$I_0(x) \sim 1$ for $x \ll 1$ and $x \in
\mathbb{R}^{+}$  (see, e.g., 9.6.7 in Ref.~\cite{abramowitz-stegun1972hom})
for the circular $\lambda(t)$ and $I_0(x) \sim e^x / \sqrt{2 \pi x}$ for $x
\gg 1$ and $x \in \mathbb{R}^+$ (see, e.g., 9.6.30 and 9.7.1 in
Ref.~\cite{abramowitz-stegun1972hom}) for the linear $\lambda(t)$---one
recovers either Eq.~\eqref{eq:circular discontinuity} or
Eq.~\eqref{eq:linear discontinuity}. Therefore,
Eq.~\eqref{eq:Fourier discontinuity} interpolates between the two limiting
cases above and can be used to accurately calculate delay times for situations
where the encircling path lies somewhere in between.

%
%
\subsection{Numerical examples}
\label{sec:Numerical examples 2}

\begin{figure}[tbf]
	\begin{center}
		\includegraphics[width=85.5mm]{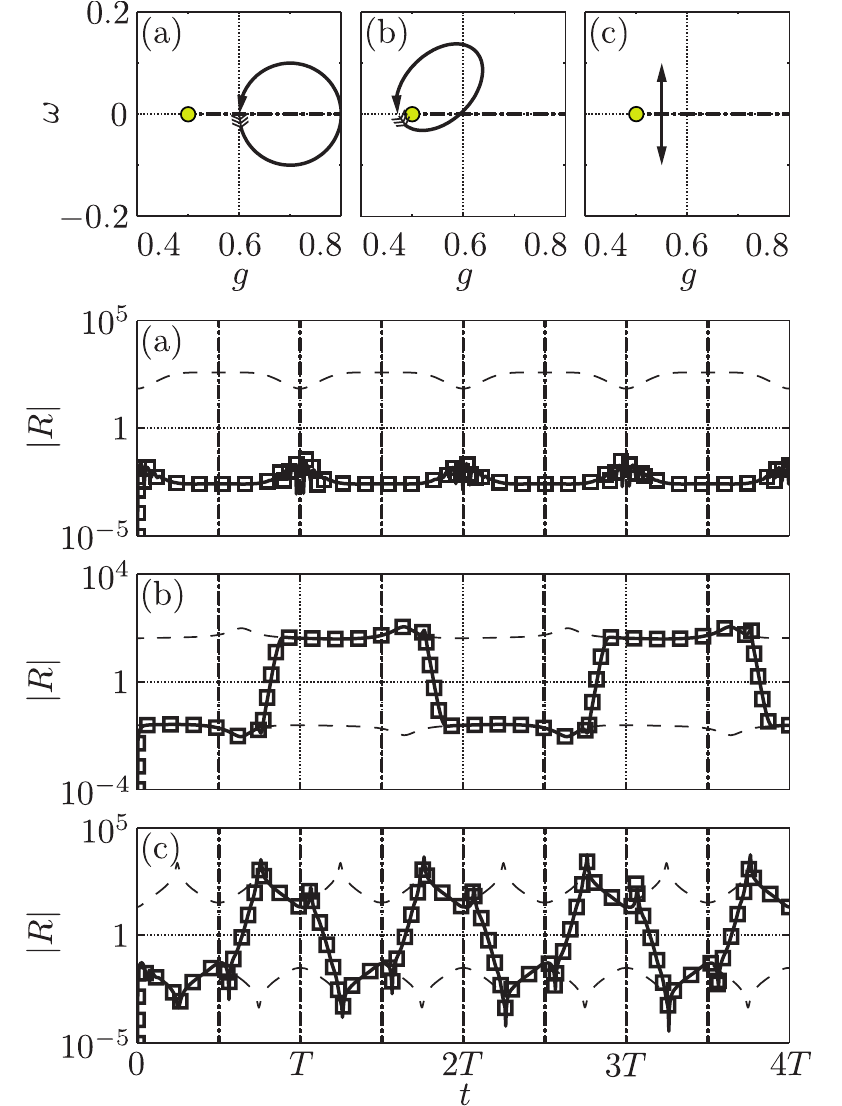}
		\caption{(Colour online.) Plots of $|R(t)|$ (three bottom panels) for
three different paths (top row). In every plot of $|R(t)|$ the solid line is
our generalised quasi-adiabatic solution, the open squares denote the
numerical solution, the dashed lines denote $\mathcal{R}^{\text{ad.}}(t)$ and
$\mathcal{R}^{\text{n.ad.}}(t)$, and the dot-dashed lines denote critical
times $t_*$. For the plots of the path, the yellow-filled circle marks the
position of the EP and the dot-dashed line is the critical line where $\Im
\lambda = 0$. The parameter settings chosen are: (a)~$r = 0.1$, $T = 200$, and
$g_{\text{o.s.}} = 0.2$; (b)~$r = 0.1$, $T = 200$, $e = 0.75$, and
$\theta_{\text{aa.}} = \pi / 4$; (c)~$L = 0.2$, $T = 200$, and
$g_{\text{o.s.}} = 0.05$ (see the main text for details on the
parametrisation).}
		\label{fig:figure4}
	\end{center}
\end{figure}

\begin{figure}[tbf]
	\begin{center}
		\includegraphics[width=85.5mm]{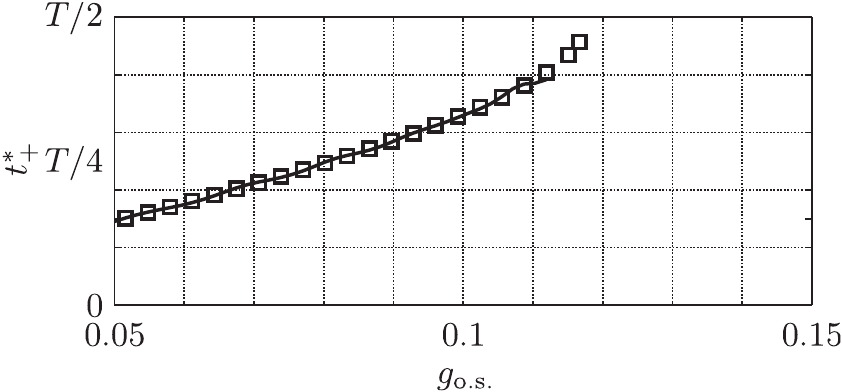}
		\caption{A plot of the departure time $t_+$ for path~(c) in
Fig.~\ref{fig:figure4} with the same parametrisation, except for
$g_{\text{o.s.}}$, which we vary. The solid line is $t_+$ as determined by
Eq.~\eqref{eq:departure time} and the open squares denote the numerically
observed departure time as determined via $|R(t)| = 1$. Good agreement is
exhibited between the analytic $t_+$s and the numerical $t_+$s. A particularly
interesting feature is that $t_+$ becomes infinite for $g_{\text{o.s.}}
\gtrsim 0.12$. For large $g_{\text{o.s.}}$ one then expects the system to
remain adiabatic for all times, as observed in Fig.~\ref{fig:figure4}(a).}
		\label{fig:figure4prime}
	\end{center}
\end{figure}

Let us now demonstrate the validity of our approach numerically for more
general examples depicted in Fig.~\ref{fig:figure4}:
\begin{enumerate}[(a)]
	\item a displaced circular path, $\omega(t) = r \sin \phi(t)$ and $g(t) =
\gamma / 2 + r \cos \phi(t) + g_{\text{o.s.}}$ where $\phi(t) = 2 \pi t / T$
and $g_\text{o.s.}$ is a variable off-set in the coupling; 
	\item a tilted elliptical path, $\omega(t) = r(t) \sin \phi(t)$ and $g(t)
= \gamma / 2 + r(t) \cos \phi(t)$ where $\phi(t) = 2 \pi t / T$, $r(t) = r_0
(1 - e^2) / \{1 + e \cos[\phi(t) + \theta_{\text{aa.}}]\}$, $e$ is the
ellipticity and $\theta_{\text{aa.}}$ is the angle of the apoapsis;
	\item oscillations along a straight path which crosses the critical line,
$\omega(t) = - L \sin \phi(t)$ and $g(t) = \gamma / 2 + g_{\text{o.s.}}$ where
$\phi(t) = 2 \pi t / T$.
\end{enumerate}
For these examples the numerically simulated solution $R(t)$ is plotted in
Fig.~\ref{fig:figure4} and compared with the generalised quasi-adiabatic
solution presented in Sec.~\ref{sec:Generalised quasi-adiabatic solution},
with $\Delta$ being evaluated numerically. We see that in all cases the 
numerical and analytic results match up perfectly. 

Let us first consider path~(a). For this path there are two dynamical
bifurcations in every period, as indicated, but the departure time as
determined by Eq.~\eqref{eq:departure time} is infinite. Therefore, the
solution never has enough time to be significantly repelled from
$R^{\text{ad.}}(t)$ before $R^{\text{ad.}}(t)$ becomes stable once again. As a
result, the system never leaves the neighbourhood of $R^{\text{ad.}}(t)$,
i.e., the solution is adiabatic. In some sense, the increased frequency of
dynamical bifurcations and the long departure time has stabilised the
adiabatic prediction. In path~(b) the opposite is the case. Here the solution
undergoes a non-adiabatic transition every period. The solution looks quite
similar to that shown in Fig.~\ref{fig:figure4}(a), except that the
non-adiabatic transitions occur earlier. One finds in this case that the delay
time is roughly $0.32 T$, slightly less than $T / 2$. Finally, in path~(c) we
have chosen again a path with two dynamical bifurcations per period, but in
this case the departure time is roughly $0.15T$, significantly less than $T /
2$. Accordingly, we observe two non-adiabatic transitions per period.

The last case can in fact be made to resemble either of the former two cases
by tuning $g_{\text{o.s.}}$, i.e., by changing the value of
$\lambda_{\Re}(t_*)$. In Fig.~\ref{fig:figure4prime} we have plotted
the departure time as a function of $g_{\text{o.s.}}$. For $g_{\text{o.s.}} <
0.05$ the quasi-adiabatic condition breaks down. For $0.05 < g_{\text{o.s.}}
\lesssim 0.12$, i.e., close to the EP, the solution resembles
Fig.~\ref{fig:figure4}(c) and we observe two non-adiabatic transitions in
every period. And around $g_{\text{o.s.}} = 0.12$ the departure time becomes
infinite, which implies that for $g_{\text{o.s.}} \gtrsim 0.12$ the dynamics
becomes fully adiabatic and resembles Fig.~\ref{fig:figure4}(a).

\section{Noise}
\label{sec:Noise}

Finally, it is important to address the influence of noise, which will be
present in any experimental implementation. To do so we simulated the dynamics
of $c_\mp$ in the presence of delta-correlated Gaussian noise $\xi(t)$ with
variance $\langle \xi(t) \xi(t') \rangle = \gamma N \delta(t - t')$. The gray
shaded area in Fig.~\ref{fig:figure3}(a) indicates the resulting distribution
of stochastic trajectories of $R(t)$ for $N = 1/10$. This resembles the case
where the initial resonator amplitude is a factor of~$10$ above the thermal
noise floor. For the first encircling period the fixed point $R^{\rm
n.ad.}(t)$ is still robust, but the delay time $t_+$ is significantly reduced.
This again demonstrates the extreme sensitivity of
Eq.~\eqref{eq:dynamical flip errors} upon initial conditions. However, the
dynamics of $R$ is self-correcting, and after the first encircling period it
settles into robust periodic dynamics much resembling the case without noise.
This surprising observation can be understood as follows. Initially, noise
causes $R$ to loose stability early. But this means that the total population
of the system is increased, and therefore the effect of the constant noise
background is reduced.

\section{Conclusion}
\label{sec:Conclusion}

In summary, we have analyzed the quasi-adiabatic evolution of non-Hermitian
systems near an EP. Our study shows that various dynamical phenomena
associated with this process can be predicted from the analysis of the
non-adiabatic transition amplitudes, $R_-(t)$ and $R_+(t)$. In particular, we
identified a characteristic switching pattern and stability loss delay. Our
analytic predictions for the delay times and the observed robustness with
respect to noise are relevant for first experimental investigations of these
effects and provide the basis for analysing similar phenomena in more complex
systems.

\emph{Acknowledgments.} The authors thank M. V. Berry, E.-M. Graefe, M.
Koller, D. O. Krimer, A. Mailybaev, A. Neishtadt, and R. Uzdin for useful
input. This work was supported by the European FP7/ITC Project SIQS (600645),
project OPSOQI (316607) of the WWTF, the Austrian Science Fund (FWF) through
SFB FOQUS F40, SFB NextLite F49-10, project GePartWave No.\@ I 1142-N27, and
the START grant Y 591-N16.

\appendix

\section{Non-standard analysis of relative non-adiabatic transition amplitues}
\label{sec:Non-standard analysis of relative non-adiabatic transition amplitues}

In this appendix we briefly summarise the motivation for the non-standard
analysis of quasi-adiabatic non-Hermitian systems~\cite{nenciu-rasche1992ota,
leclerc-jolicard2012tro, ibenez-muga2014acf} and its application to relative
non-adiabatic transition amplitudes~\cite{neishtadt1987pos, neishtadt1988pos,
diener-diener1991md, berglund-gentz2006nip, bohun2011sld}. In order to
facilitate a simple but rigorous mathematical treatment we augment the
notation of the paper by introducing a dimensionless time
\begin{equation}
	s := \frac{t}{T} ,
\end{equation}
where $T^{-1}$ is considered `small', and rewrite the governing equation of
motion
\begin{equation}
	\dot{\mathcal{U}} = - i \textbf{M}(s) \mathcal{U}
\end{equation}
with $\mathcal{U}(0) = 1$ and where $\mathcal{U}$ is the evolution operator
and the dot denotes the derivative with respect to time $t$ as usual.

\subsection{Quasi-adiabaticity}
\label{sec:Quasi-adiabaticity}

We assume $\textbf{M}(s)$ to be diagonalisable with eigenvalues $\lambda_i(s)$
for all $s$. Since $\textbf{M}(s)$ is non-Hermitian it does not in general
have an orthonormal eigenbasis in the sense of Dirac but rather a biorthogonal
eigenbasis: a set of right eigenvectors $\vec{r}_i(s)$ defined via
$\textbf{M}(s) \vec{r}_i(s) = \lambda_i(s) \vec{r}_i(s)$ and a set of left
eigenvectors $\vec{l}_i^T(s)$ defined via $\vec{l}_i^T(s) \textbf{M}(s) =
\vec{l}_i^T(s) \textbf{M}(s)$ such that $\vec{l}_i^T(s) \vec{r}_j(s) =
\delta_{i, j}$. Ideal adiabatic dynamics may be defined as that for which the
dynamical coefficients of the instantaneous eigenvectors decouple. In a
parallel transported eigenbasis, i.e., $\vec{l}_i^T(s) \vec{r}_i'(s) = 0$
where the prime denotes the derivative with respect to $s$, the adiabatic
solution, or adiabatic prediction, is
\begin{equation}\label{eq:adiabatic solution}
	\mathcal{U}_{i, j}(t) = \delta_{i, j}  
\end{equation}
where we have expanded the evolution operator $\mathcal{U}$ thus
\begin{equation}
	\mathcal{U}(t) = \sum_{i, j} \mathcal{U}_{i, j} e^{- i T \int_0^s ds' \lambda_i(s')} \vec{r}_i(s) \vec{l}_j^T(0) .
\end{equation}

In the adiabatic solution the interaction between the dynamical coefficients
of the instantaneous eigenvectors due to the finite variation of these
eigenvectors is ignored. The full equation of motion for $\mathcal{U}$
expanded as above is
\begin{equation}
	\dot{\mathcal{U}}_{p, q} = - i \sum_{i \neq p} T^{-1} \tilde{f}_{p, i}(s)
		e^{- i T \int_0^s ds' [\lambda_i(s') - \lambda_p(s')]} \mathcal{U}_{i, p}
\end{equation}
where we have defined
\begin{equation}
	\tilde{f}_{p, i}(s) := - i \vec{l}_p^T(s) \vec{r}_i'(s) .
\end{equation}
The adiabatic solution ignores $\tilde{f}_{p, q}(s)$ for $p \neq q$. Assuming
the system to be initialised to instantaneous eigenvector $q$, first order
perturbation theory yields that the solution for the coefficient $x_p$ of
instantaneous eigenvector $p$ where $p \neq q$ is
\begin{equation}
	x_p(t) \simeq \frac{T^{-1}
		\tilde{f}_{p, q}(s)
		e^{- i T \int_0^{s} ds' [\lambda_q(s') - \lambda_p(s')]}
		}{\lambda_q(s) - \lambda_p(s)} .
\end{equation}
This expression vanishes linearly with $|T^{-1} / [\lambda_q(s) -
\lambda_p(s)]$ but diverges exponentially with $T \Im \int_0^s ds'
[\lambda_q(s') - \lambda_p(s')]$ if $\int_0^s ds' \lambda_q(s') > \int_0^s ds'
\lambda_p(s')$. Second order perturbation theory contains no more information
as regards $x_p$ but does reveal that $x_q(t)$ differs from unity with an
analogous scaling. The traditional quantum adiabatic condition
\begin{equation}\label{eq:quasi-adiabatic condition}
	\varepsilon_{p, q}(t) := \left|
		\frac{T^{-1} \tilde{f}_{p, q}(s)}{\lambda_q(s) - \lambda_p(s)}
		\right| \ll 1
\end{equation}
therefore only ensures adiabaticity for those elements of $\mathcal{U}$ for
which $\Im \int_0^s ds' \lambda_i(s')$ is greatest, i.e., the least
dissipative instantaneous eigenvectors. It obviously cannot be the case that
every eigenvector is least dissipative, unless all are degenerate, and it is
therefore impossible that the adiabatic solution
Eq.~\eqref{eq:adiabatic solution} hold. In the context of non-Hermitian
systems it therefore seems pertinent to call
Eq.~\eqref{eq:quasi-adiabatic condition} the \emph{quasi}-adiabatic condition.
So long as we initialise to the least dissipative instantaneous eigenstate and
so long as this eigenstate remains the least dissipative, the quasi-adiabatic
condition ensures adiabaticity. But if we initialise to an eigenstate that is
not the least dissipative, or the quality of being least dissipative is
exchanged, then perturbation theory breaks down.

\subsection{Relative non-adiabatic transition amplitudes as a slow-fast system}
\label{sec:Relative non-adiabatic transition amplitudes as a slow-fast system}

In the main text we argued that for our two-dimensional case the simplest
encompasing dynamical description of adiabaticity is afforded by the relative
non-adiabatic transition amplitudes---the ratios of the elements of the
evolution operator expressed in a parallel transported eigenbasis. Let us
recall the equation of motion for the relative non-adiabatic transition
amplitude $R(t)$ as defined in the paper:
\begin{equation}
	\dot{R} = - 2 i \lambda(s) R - i T^{-1} \tilde{f}(s) (1 + R^2) .
\end{equation}
Treating $\lambda$ and $\tilde{f}$ as dynamical variables themselves, in the
limit $T^{-1} \rightarrow 0$ this becomes
\begin{equation}\label{eq:fast time-scale}
	\dot{R} = - 2 i \lambda(s_0) R - i T^{-1} \tilde{f}(s_0) (1 + R^2)
\end{equation}
where $s_0 = T^{-1} t_0$ and $t_0$ is the initial time. On the other hand, we
may rewrite the equation of motion using $s$ as the independent variable,
\begin{equation}
	T^{-1} R' = - 2 i \lambda(s) R - i T^{-1} \tilde{f}(s) (1 + R^2) ,
\end{equation}
whereupon similarly taking $T^{-1} \rightarrow 0$ yields
\begin{equation}\label{eq:slow time-scale}
	0 = - 2 i \lambda(s) R - i T^{-1} \tilde{f}(s) (1 + R^2) .
\end{equation}
The difference between Eqs.~\eqref{eq:fast time-scale}
and~\eqref{eq:slow time-scale} is that the former is over a time-scale of
order $T^{-1}$ and is hence \emph{fast}, whilst the latter is over a
time-scale of order $1$ and is hence \emph{slow}; we have a slow-fast system.
Furthermore, we notice here that the fast time-scale equation is differential
whilst the slow algebraic. This is often taken as the definition of a
singularly perturbed system and it means that any perturbative approach in
$T^{-1}$ can only be valid for times of order $T^{-1}$. In order to study the
long time behaviour we must turn to non-standard analysis.

\subsection{Slow manifolds}
\label{sec:Slow manifolds}

The solutions of the slow time-scale Eq.~\eqref{eq:slow time-scale} are the
fixed points of the fast time-scale Eq.~\eqref{eq:fast time-scale} and as such
are known as \emph{instantaneous} fixed points. We recall their approximate
expressions from the main text:
\begin{gather}
	\begin{aligned}
		R^{\text{ad.}}(s) &\simeq - \frac{T^{-1} \tilde{f}(s)}{2 \lambda(s)} \text{ and}\\
		R^{\text{n.ad.}}(s) &\simeq - \frac{2 \lambda(s)}{T^{-1} \tilde{f}(s)} .
	\end{aligned}
\end{gather}
Since these are the fixed points of Eq.~\eqref{eq:fast time-scale} we may use
Eq.~\eqref{eq:fast time-scale} to perform a stability analysis. One finds that
$R^{\text{ad.}}(s)$ is stable if and only if $\Im \lambda(s) < 0$, whilst
$R^{\text{n.ad.}}(s)$ is stable if and only if $\Im \lambda(s) > 0$, and the
possible local phase portraits are stable star, stable spiral, centre,
unstable spiral, and unstable star~\cite{glendinning1994sia}. Evidently, the
only possible bifurcation is from a stable spiral to an unstable spiral
through a centre. The locus of points $R^{\text{ad.}}(s)$ over $s$ is called
a slow manifold and denoted $M^{\text{ad.}} = \{R^{\text{ad.}}(s) : s\}$.
Similarly for $R^{\text{n.ad.}}(s)$.

\subsection{Adiabatic manifolds}
\label{sec:Adiabatic manifolds}

Due to finite variations in $\lambda(s)$ and $\tilde{f}(s)$ the slow manifolds
$M^{\text{ad.}}$ and $M^{\text{n.ad.}}$ are in fact not locally invariant.
Nevertheless, a theorem due to N. Fenichel~\cite{fenichel1979gsp} ensures the
existence of locally invariant manifolds in a $T^{-1}$-neighbourhood of
$M^{\text{ad.}}$ and $M^{\text{n.ad.}}$. These locally invariant manifolds are
called adiabatic manifolds and denoted $\mathcal{M}^{\text{ad.}}$ and
$\mathcal{M}^{\text{n.ad.}}$ respectively. The adiabatic manifolds do not obey
a simple equation of motion, but we may find a good approximation by
considering the particular integral of the $N$-times linearised equation of
motion. We focus on $\mathcal{M}^{\text{ad.}}$ for clarity. Following the
argument of example~2.1.10 in Ref.~\cite{berglund-gentz2006nip} one arrives at
$\mathcal{M}^{\text{ad.}} = \{\mathcal{R}^{\text{ad.}}(s) : s\}$ where
\begin{equation}\label{eq:adiabatic manifold}
	\mathcal{R}^{\text{ad.}}(s) \simeq \sum_{n = 0}^{N - 1} T^{-n} \left(
		\frac{-1}{2 i \lambda(s)} \frac{d}{ds}
		\right)^n R^{\text{ad.}}(s)
\end{equation}
and $N$ is an optimal truncation with a remainder of order $e^{- C / T^{-1}}$
for some $C > 0$. The expression for $\mathcal{R}^{\text{n.ad.}}(s)$ is
analogous. We describe $\mathcal{R}^{\text{ad.}}(s)$ and
$\mathcal{R}^{\text{n.ad.}}(s)$ as attractive or unattractive analogously to
$R^{\text{ad.}}(s)$ and $R^{\text{n.ad.}}(s)$ being stable or unstable
respectively.

\subsection{Stability loss delay}
\label{sec:Stability loss delay}

At certain critical times $t_*$, or $s_*$, the stability of the instantaneous
fixed points swaps. For example, $R^{\text{ad.}}(s)$ becomes unstable and
$R^{\text{n.ad.}}(s)$ becomes stable at $s = s_*$ such that $\Im \lambda(s_*)
= 0$ and $\Im \lambda'(s_*) > 0$. One might na{\"i}vely suppose an immediate
transition between $M^{\text{ad.}}$ and $M^{\text{n.ad.}}$ at $s = s_*$, but
this is not the case. The bifurcation is dynamical and the type is degenerate
Hopf, which in general exhibits the phenomenon known as \emph{stability loss
delay}: the solution $R(t)$ continues to follow, say, $R^{\text{ad.}}(s)$ for
a significant time past its loss of stability. Following the argument in
Sec.~2 of Ref.~\cite{diener-diener1991md} one finds that away from $s = s_*$
the solution has the asymptotic expansion
\begin{equation}
	R(t) \sim A e^{\tilde{\Psi}(s) / T^{-1}} + \mathcal{R}^{\text{ad.}}(s)
\end{equation}
where
\begin{equation}
	\tilde{\Psi}(s) = - 2 i \int_{s_*}^s ds' \lambda(s') ,
\end{equation}
whilst at $s = s_*$ the solution exhibits the discontinuity
\begin{equation}
	\Delta = - i \int_{s_-^*}^{s_+^*} ds \tilde{f}(s) e^{- \tilde{\Psi}(s) / T^{-1}}
\end{equation}
where $s_-^* < s_*$ and $s_+^* > s_*$ are the intersections of the level curve
of $\Re \Psi$ that includes the point $z_* \in \mathbb{C}$ such that
$\lambda(z_*) = 0$. In order to incorporate this discontinuity in the
asymptotic expansion of the solution for $R(t)$ we add the term $\Theta(s -
s_*) \Delta e^{\tilde{\Psi}(s) / T^{-1}}$ where $\Theta$ is the Heaviside step
function. This term is proportional to $e^{(\tilde{\Psi}(s) -
\tilde{\Psi}(z_*))/ T^{-1}}$, and therefore diverges as $T^{-1} \rightarrow 0$
for any time $t$ such that $\Re \tilde{\Psi}(s) - \Re \tilde{\Psi}(z_*) > 0$.
Thus, the times $t_-^* < t_*$ and $t_+^* > t_*$ corresponding to $s_-^*$ and
$s_+^*$ are such that:
\begin{enumerate}[(i)]
	\item if the solution enter a neighbourhood of $M^{\text{ad.}}$ before
$t_-^*$ then it must leave at $t_+^*$;
	\item if the solution enter a neighbourhood of $M^{\text{ad.}}$ after
$t_-^*$ at $t_- < t_*$ then it must leave at $t_+ > t_*$ such that $\Re
\tilde{\Psi}(s_+) - \Re \tilde{\Psi}(s_-) = 0$;
	\item if the solution leave a neighbourhood of $M^{\text{ad.}}$ after
$t_+^*$ then it must have entered at $t_-^*$.
\end{enumerate}
Since the third case is sure to be rare, one typically calls $t_+^*$ the
\emph{maximal delay time}. Note that, this maximal delay time is precisely the
quasi-adiabatic limit of the delay time $t_+$ calculated via
Eq.~\eqref{eq:departure time}.

For an analytic example of a maximal delay time, let us consider the path
analysed in Sec.~\ref{sec:Fixed points and stability loss delay}. In this case
the level curves of $\Re \tilde{\Psi}$ are
\begin{equation}
	e^{- \pi \Im z} \cos(\pi \Re z) \simeq \text{const} .
\end{equation}
Evidently, any $s_- < 0$ such that $s_- > - 1 / 2$ is connected to $s_+ = -
s_- > 0$ by these level curves. The maximal delay time is therefore $t_+^* = T
/ 2$, which agrees with the main text.

\begin{figure}[tbf]
	\begin{center}
		\includegraphics[width=81.468mm]{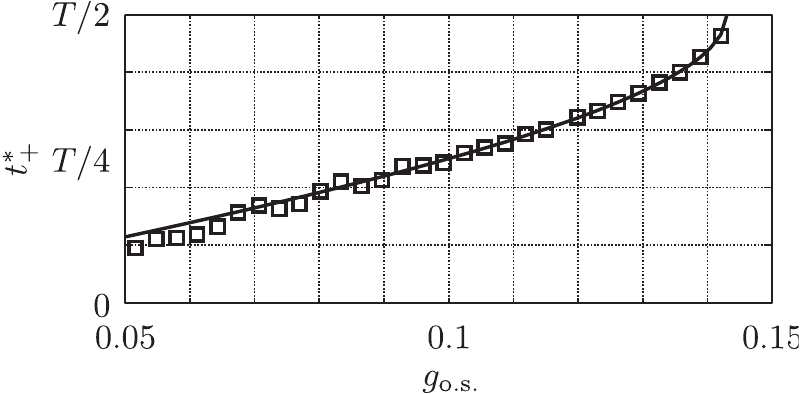}
		\caption{Plot of the theoretical maximal delay time (solid line) and
numerically observed departure times from a $5\%$-neighbourhood of
$R^{\text{ad.}}(s)$ (open squares) for case~(c) from
Sec.~\ref{sec:Numerical examples 2}.}
		\label{fig:appx_max_delay}
	\end{center}
\end{figure}
 
For a numerical example, let us consider case~(c) from
Sec.~\ref{sec:Numerical examples 2}. Here we calculate the theoretical maximal
delay time by numerically finding the complex root of $\lambda(z)$, i.e.,
$z_*$, and then numerically finding where the level curve of $\Re
\tilde{\Psi}$ on which $z_*$ lies intersects the real axis. The results are
plotted in Fig.~\ref{fig:appx_max_delay}. Also plotted in the figure are the
results of a numerical solution where we have initialised to a neighbourhood
of $R^{\text{ad.}}(s)$ at $s = s_-^*$ and asked for when the numerical
solution departs from this neighbourhood. The agreement between the
theoretical maximal delay time and the numerically observed departure time is
very good. Furthermore, we see qualitatively the same results as those shown
in Fig.~\ref{fig:figure4prime}; the quantitative difference is simply due to
the finite time required for $R(t)$ to leave a small neighbourhood of
$M^{\text{ad.}}$ and reach $|R(t)| = 1$.

%
\section{Non-adiabatic transitions as a manifestation of the Stokes phenomenon of asymptotics}
\label{sec:Non-adiabatic transitions as a manifestation of the Stokes phenomenon of asymptotics}

The Stokes phenomenon of asymptotics is that subdominant exponentials in the
asymptotic expansion of certain functions disappear and reappear in different
sections of the complex plane with different coefficients. In Stokes'
words~\cite{stokes1902otd},
\begin{quotation}
\dots the inferior term enters as it were into the mist, is hidden for a
little while from view, and comes out with its coefficient changed.
\end{quotation}
M. V. Berry and R. Uzdin~\cite{berry-uzdin2011snh} uncovered the presence of
the Stokes phenomenon of asymptotics in the solutions of specific exactly
solvable models of quasi-adiabatic non-Hermitian systems and identify
non-adiabatic transitions in such systems as a manifestation. In this appendix
we review one such example.

Let us consider again the parametrisation studied in
Sec.~\ref{sec:Fixed points and stability loss delay}: $\omega(t) = r \sin
\phi(t)$, $\gamma(t) = 1$, and $g(t) = 1/2 + r \cos \phi(t)$ with $r \ll 1$
and $\dot{\phi}(t) = \dot{\phi} = 2 \pi / T = \text{const}$. An exact solution
for this example is presented by M. V. Berry~\cite{berry2011ope} which we now
review. Constructing the analagous quantity to $R$ in the circular basis
\begin{equation}\label{eq:circular basis}
	\vec{r}_\circlearrowright = \frac{1}{\sqrt{2}} \begin{pmatrix}
		1 \\ i
		\end{pmatrix} \text{ and } \vec{r}_\circlearrowleft = \frac{1}{\sqrt{2}} \begin{pmatrix}
		1 \\ -i
		\end{pmatrix} ,
\end{equation}
which we denote $p$, i.e., $\vec{x}(t) = c_\circlearrowright(t)
\vec{r}_\circlearrowright + c_\circlearrowleft(t) \vec{r}_\circlearrowleft$
and $p(t) = c_\circlearrowleft(t) / c_\circlearrowright(t)$, one arrives at
the equation of motion
\begin{equation}
	\dot{p} \simeq r e^{i \phi(t)} + p^2 ,
\end{equation}
where we have used $r \ll 1$. Note that, $\vec{r}_\circlearrowright$ and
$\vec{r}_\circlearrowleft$ are not eigenvectors. Introducing the new
independent variable $\zeta = (i / 2) (2 \varepsilon)^{-1} e^{i \phi(t) / 2}$
where $\varepsilon = \pi / (4 \sqrt{r} T)$ and using the ansatz $p(t) = - (d /
dt) \log f(\zeta)$ yields
\begin{equation}
	\zeta^2 \frac{d^2 f}{d \zeta^2} + \zeta \frac{d f}{d \zeta} + \zeta^2 f = 0 .
\end{equation}
This is the zeroth order Bessel equation and the solution from time $t = t_0$
is thus
\begin{equation}
	p(t) = \frac{i}{2} \dot{\phi} \zeta \frac{\mathcal{C}_1(\zeta)}{\mathcal{C}_0(\zeta)}
\end{equation}
where $\mathcal{C}_n(\zeta) = c_J J_n(\zeta) + c_Y Y_n(\zeta)$ is a linear
combination of order~$n$ Bessel functions of the first and second kind with
the ratio
\begin{equation}
	\frac{c_Y}{c_J} = - \frac{2 i p(t_0) J_0(\zeta_0) + \dot{\phi} \zeta_0 J_1(\zeta_0)}
		{2 i p(t_0) Y_0(\zeta_0) + \dot{\phi} \zeta_0 Y_1(\zeta_0)}
\end{equation}
where $p(t_0)$ is the initial condition for $p$ and $\zeta_0 = (i / 2) (2
\varepsilon)^{-1} e^{i \phi(t_0) / 2}$.

In order to translate this result into an expression for $R(t)$ we have only
to transform from the circular basis Eq.~\eqref{eq:circular basis} to the
original basis and then from that to the parallel transported eigenbasis
Eq.~\eqref{eq:eigenvectors},
\begin{equation}
	\vec{r}_-(t) = \begin{pmatrix}
		\cos \vartheta(t) / 2 \\ \sin \vartheta(t) / 2
		\end{pmatrix} \text{ and }
		\vec{r}_+(t) = \begin{pmatrix}
		- \sin \vartheta(t) / 2 \\ \cos \vartheta(t) / 2
		\end{pmatrix}
\end{equation}
where $\tan \vartheta(t) = - g(t) / [\omega(t) + i \gamma(t) / 2]$.
Recognising the effect of such transformations on $p$ and $R$ as M{\"o}bius
transformations, it is immediately seen that
\begin{equation}
	p(t_0) = e^{i \vartheta(t_0)} \frac{1 + i R(t_0)}{1 - i R(t_0)} ,
\end{equation}
where $R(t_0)$ is the initial condition for $R$, and
\begin{equation}
	R(t) = i \frac{1 - e^{- i \vartheta(t)} p(t)}{1 + e^{- i \vartheta(t)} p(t)} .
\end{equation}

Let us focus on the asymptotic expansion about $t = t_*$ where
$R^{\text{ad.}}(t)$ becomes unstable. Recalling that $\lambda(t) \simeq
\sqrt{r} e^{i \phi(t) / 2}$ and $\Im \lambda(t_*) = 0$ we see that with
$\zeta_* = (i / 2) (2 \varepsilon)^{-1} e^{i \phi(t_*) / 2}$ we have $\Re
\zeta_* = 0$. Without loss of generality, we suppose $\Im \zeta_* > 0$. Using
9.1.35, 9.1.36, 9.2.1, and 9.2.2 in Ref.~\cite{abramowitz-stegun1972hom}
and assuming that the solution is exponentially close to $R^{\text{ad.}}(t)$
by $t = t_*$ yields
\begin{widetext}
\begin{equation}\label{eq:Berrys asymptotic solution}
	R(t) \sim i \frac{
		1 - e^{\vartheta(t)} p^{\text{ad.}}(t)
		+ 2 i \Theta(- \Re \zeta) e^{2 i \zeta} [1 - e^{\vartheta(t)} p^{\text{n.ad.}}(t)]
		}{
		1 + e^{\vartheta(t)} p^{\text{ad.}}(t)
		+ 2 i \Theta(- \Re \zeta) e^{2 i \zeta} [1 + e^{\vartheta(t)} p^{\text{n.ad.}}(t)]
		}
\end{equation}
\end{widetext}
for $- \pi / 2 < \arg \zeta < 3 \pi / 2$ where $p^{\text{ad.}}(t)$ and
$p^{\text{n.ad.}}(t)$ correspond to $R^{\text{ad.}}(t)$ and
$R^{\text{n.ad.}}(t)$ respectively and $\Theta$ is the Heaviside step
function. The discontinuity in this asymptotic expansion is precisely the
Stokes phenomenon of asymptotics. For $t \in (t_* - T / 2, t_* + T / 2)$ the
discontinuous term is subdominant and $R(t) \sim R^{\text{ad.}}(t)$, whereas
for $t > t_* + T / 2$ it is dominant and $R(t) \sim R^{\text{n.ad.}}(t)$. The
connection to the expansion Eq.~\eqref{eq:asymptotic solution} from
Sec.~\ref{sec:Fixed points and stability loss delay} is more clearly seen by
expanding to first order about $R^{\text{ad.}}(t)$:
\begin{equation}
	R(t) \approx R^{\text{ad.}}(t) - 2 \Theta(t - t_*) e^{- \exp(i \phi(t) / 2) / (2 \varepsilon)} .
\end{equation}
Comparing this expansion to Eq.~\eqref{eq:asymptotic solution} we find very
good agreement. The small difference that here the discontinuity at $t = t_*$
is $\Delta = - 2 e^{- 1 / (2 \varepsilon)}$ whereas in
Sec.~\ref{sec:Analytic examples} we found $\Delta = - \pi e^{- 1 / (2
\varepsilon)}$ is principally due to the asymptotic expansions employed in
calculating Eq.~\eqref{eq:Berrys asymptotic solution}, which differ from those
used in calculating Eq.~\eqref{eq:asymptotic solution}.

\end{document}